\documentclass[fleqn,usenatbib]{mnras}

\usepackage{newtxtext,newtxmath}

\usepackage[T1]{fontenc}

\DeclareRobustCommand{\VAN}[3]{#2}
\let\VANthebibliography\thebibliography
\def\thebibliography{\DeclareRobustCommand{\VAN}[3]{##3}\VANthebibliography}

\usepackage{graphicx}	
\usepackage{amsmath}	
\usepackage{amssymb}	

\title[ultra-low-frequency GWs from an eccentric SMBH binary]{Constraints on ultra-low-frequency gravitational waves from an eccentric supermassive black hole binary}

\author[Kikunaga et al.]{
Tomonosuke Kikunaga,$^{1}$\thanks{E-mail: amqmysuto@gmail.com}
Shinnosuke Hisano$^{1}$
Hiroki Kumamoto,$^{1}$
and Keitaro Takahashi$^{1,2,3}$
\\
$^{1}$Kumamoto University, Graduate School of Science and Technology, Kumamoto, 860-8555, Japan\\
$^{2}$Kumamoto University, International Research Organization for Advanced Science and Technology, Kumamoto, 860-8555, Japan\\
$^{3}$National Astronomical Observatory of Japan, 2-21-1 Osawa, Mitaka, Tokyo 181-8588, Japan
}

\date{Accepted XXX. Received YYY; in original form ZZZ}

\pubyear{2021}

\begin{document}
\label{firstpage}
\pagerange{\pageref{firstpage}--\pageref{lastpage}}
\maketitle

\begin{abstract}
Milli-second pulsars with highly stable periods can be considered as very precise clocks and can be used for pulsar timing array (PTA) which attempts to detect nanoheltz gravitational waves (GWs) directly. Main sources of nanoheltz GWs are supermassive black hole (SMBH) binaries which have sub-pc-scale orbits. On the other hand, a SMBH binary which is in an earlier phase and has pc-scale orbit emits ultra-low-frequency ($\lesssim 10^{-9}\,\mathrm{Hz}$) GWs cannot be detected with the conventional methodology of PTA. Such binaries tend to obtain high eccentricity, possibly $\sim 0.9$.  In this paper, we develop a formalism for extending constraints on GW amplitudes from single sources obtained by PTA toward ultra-low frequencies considering the waveform expected from an eccentric SMBH binary. GWs from an eccentric binaries are contributed from higher harmonics and, therefore, have a different waveform those from a circular binary. Furthermore, we apply our formalism to several hypothetical SMBH binaries at the center of nearby galaxies, including M87, using the constraints from NANOGrav's 11-year data set. For a hypothetical SMBH binary at the center of M87, the typical upper limit on the mass ratio is $0.16$ for eccentricity of $0.9$ and semi-major axis of $a=1~\mathrm{pc}$, assuming the binary phase to be the pericenter.
\end{abstract}




\section{Introduction}

Milli-second pulsars (MSPs) with very stable periods can be used as precise clocks. If gravitational waves (GWs) exist in the space between the earth and pulsars, the arrival time of pulses is changed. With this effect, we can detect low-frequency GWs ($10^{-9}$ - $10^{-6}$ Hz) and this method is called pulsar timing array (PTA) \citep{Foster}. So far, three PTA experiments have been conducting long-term observations of MSPs: the Parkes PTA in Australia \citep{ParkesPTA}, the European PTA \citep{EuropeanPTA}, and NANOGrav in North America \citep{NANOGrav}. Further, Chinese PTA \citep{CPTA} and Indian PTA \citep{InPTA} have started in recent years.

One of the major GW sources in the frequency range of PTA is supermassive black hole (SMBH) binaries in the late stage of the evolution with sub-pc scale orbital radii. NANOGrav have released 11 years of pulsar observation data \citep{NANOGrav2018} and searched GWs from an individual source \citep{NANOGrav11year}. Although they could not find GWs in their 11-year data set they placed 95\% upper limits on GWs amplitude and a chirp mass of a hypothetical SMBH binary in the Virgo Cluster. Recently, they also put limits on mass of SMBH binary in nearby massive galaxies \citep{NANOGrav2021}.

On the other hand, binaries in the early stage of the evolution interact efficiently with the environmental gas and stars and their orbital radii are reduced rapidly. However, when the orbital radius becomes a few pc, the interaction becomes weak and the orbital radius shrink only through GW emission. GW emission at this stage is not efficient and the expected merger time exceeds the Hubble time \citep{FPB1, FPB2}. This is called "the final parsec problem". Therefore, to understand the evolution of SMBH binaries, it is important to detect GWs from binaries at this stage. However, such GWs have sub-nHz frequencies and are out of the sensitivity range of the conventional PTA method. 

In our previous work \citep{Yonemaru2016}, we proposed a new detection method for these ultra-low-frequency GWs from a single source. The method utilizes the fact that the spin-down rate of MSPs is biased by ultra-low-frequency GWs and it was shown that the time derivative of GW amplitude is constrained from the statistics of spatial pattern of pulsar spin-down rates in the sky. Then we evaluated the sensitivity with Monte-Carlo simulations \citep{Yonemaru2018, Hisano2019} and put constraints on GWs from the Galactic Center and M87 as $\dot{h} < 6.2 \times 10^{-18}\,\mathrm{sec}^{-1}$ and $\dot{h} < 8.1 \times 10^{-18}\,\mathrm{sec}^{-1}$,  
respectively, for $f_{\mathrm{GW}}=1/(1000\,\mathrm{year})$ \citep{Kumamoto2019}, where $h$ is the GW amplitude and the dot represents the time derivative.

\defcitealias{MCJ15}{MCJ15}
On the other hand, in \citet{MCJ15} (hereafter \citetalias{MCJ15}), they extended the sensitivity curve of PTAs toward lower frequencies in a different way. They considered the Taylor expansion of GW waveform in low-frequency limit and proposed to extract the GW amplitude from the third and higher order terms, while terms below the second-order are absorbed by pulsar parameters. Then, signal-to-noise ratio of GWs in lower frequencies were calculated. As a result, the sensitivity curve of GWs was shown to be proportional to $f^{-2}$ at lower frequencies.

A critical assumption in \citetalias{MCJ15} is that an SMBH binary has a circular orbit. Therefore, their method is not applicable to binaries with eccentric orbits because GWs from an eccentric binary include higher harmonics and, therefore, have a very different waveform compared to that of GWs from a circular binary \citep{Peters1963}. In fact, it has been shown by numerical simulations that pc-scale SMBH binaries tend to obtain high eccentricity (typically $ e = 0.9 $ for mass ratio $q \sim 10^{-3}$) via interaction with their environment \citep{Sesana2010}. Thus, it is important to probe sub-nHz GWs from not only circular binaries but eccentric binaries. In this paper, we propose a method which is applicable to eccentric SMBH binaries extending the formalism of \citetalias{MCJ15}.

The structure of this paper is following. In section \ref{section2}, we briefly review the Kepler problem and analytical solution of GWs from an eccentric binary. Then, upper limits on eccentric GWs amplitude are derived expanding the \citetalias{MCJ15}'s method in section \ref{section3}. In section \ref{section4}, we apply our formalism to several possible SMBH binaries in nearby galaxies and derive limits on binary parameters. Finally, our results are summarized in section \ref{section5}. For the rest of this paper we set $c=G=1$, unless otherwise specified.

\section{Eccentric gravitational waveform}
\label{section2}

\subsection{Eccentric SMBH binary}

Let us consider an eccentric binary system consisting of masses $m_1$ and $m_2$ ($m_1>m_2$), reiterating some of the notation and formalism of \cite{Yunes2009} and \cite{Taylor2016}. Such a system is well-known as the Kepler problem. Considering a coordinate system with a total mass $M_{\mathrm{tot}}$ as the center of mass, the binary system can be described as
\begin{align}
    r&=a(1-e\cos u), \\
    \omega(t-t_0)&=l=u-e\sin u, \\
    \Phi -\Phi_0&=2\arctan\left[\left(\frac{1+e}{1-e}\right)^{1/2}\tan\frac{u}{2} \right], \\
    \omega &= 2\pi f = 2\pi \sqrt{\frac{M_{\mathrm{tot}}}{a^3}},
    \label{eq:f}
\end{align}
where $r$ is the distance from $m_1$ to $m_2$, $a$ is the semi-major axis of the orbit, $e$ is the orbital eccentricity, $u$ is the eccentric anomaly, $\omega$ is the average angular frequency, $l=\omega t +l_0=2\pi f t + l_0$ is the mean anomaly, $\Phi$ is the orbital phase, and $\Phi_0=\Phi(0)$. In order to express $\Phi$ as the function of time, we use the first Bessel function $J_n$ and we have,
\begin{align}
    \cos \Phi&=-e+\frac{2}{e}(1-e^2)\sum_{n=1}^\infty J_n(ne)\cos(nl), \\
    \sin \Phi&=(1-e^2)^{1/2}\sum_{n=1}^\infty [J_{n-1}(ne)-J_{n+1}(ne)]\sin(nl).
\end{align}

\subsection{GW waveform}

Imposing the transverse-traceless gauge (TT gauge), the GW tensor can be expressed as a superposition of two polarization modes and given by,
\begin{equation}
    h_{ij}(t, \hat{\Omega})=h_+(t)e^+_{ij}(\hat{\Omega})+h_{\times}(t)e^\times _{ij}(\hat{\Omega}),
\end{equation}
where $\hat{\Omega}$ is the direction of GW propagation, and $e_{ij}^{+,\times}$ are polarization tensors. If a SMBH binary has non-zero eccentricity, GWs emitted from it have higher harmonics components and the 
amplitude of two polarization modes is as follows:
\begin{align}
    h_{+}(t)=& h_0\sum_{n}-\left(1+\cos ^{2} \iota\right)\left[a_{n}(t) \cos (2 \gamma)-b_{n}(t) \sin (2 \gamma)\right] \nonumber \\
    &+\left(1-\cos ^{2} \iota\right) c_{n}(t), \\ 
    h_{\times}(t)=& h_0\sum_{n} 2 \cos \iota\left[b_{n}(t) \cos (2 \gamma)+a_{n}(t) \sin (2 \gamma)\right], 
\end{align}
where
\begin{align}
    h_0=& \frac{2m_1m_2}{Da},\label{eq:h0}\\
    a_{n}(t)=&x_{a_n}(e)\cos[nl(t)], \label{eq:a_n}\\
    b_{n}(t)=&x_{b_n}(e)\sin[nl(t)], \\
    c_{n}(t)=&x_{c_n}(e)\cos[nl(t)], \\
    x_{a_n}(e) =&-\frac{n}{2}\left[J_{n-2}(n e)-2 e J_{n-1}(n e)+\frac{2}{n}J_{n}(n e)\right. \nonumber\\ &\left. +2 e J_{n+1}(n e)-J_{n+2}(n e)\right], \\
    x_{b_n}(e) =&-\frac{n}{2}\sqrt{1-e^{2}}\left[J_{n-2}(n e)-2 J_{n}(n e)+J_{n+2}(n e)\right], \\
    x_{c_n}(e) =&J_n(ne).\label{eq:x_c_n}
\end{align}
Here, $m_1$ and $m_2$ are the mass of the main SMBH and second BH, respectively, $D$ is the distance from the Earth to the source, $\iota$ is the orbital inclination, and $\gamma$ is the azimuthal angle measuring the direction of pericenter. In the case of a circular binary, i.e. $e=0$, only $n=2$ terms remain in Eqs. (\ref{eq:a_n}) to (\ref{eq:x_c_n}). In this expression, $h_{+,\times}$ depends on time through the trigonometric functions. Then, we combine them into a cosine function:
\begin{align}
    h_M(t) &= h_0 \sum_n \sqrt{A_{M,n}^2 + B_{M,n}^2}\cos \left(nl(t) + \alpha_{M,n} \right) \label{eq:GWsHarmonics}\\
    A_{+,n} &= \left( 1 + \cos^2 \iota\right)x_{b_n} \sin(2\gamma) \\
    B_{+,n} &= \left(1 - \cos^2 \iota\right)x_{c_n} - \left(1+\cos^2 \iota\right)x_{a_n}\cos(2\gamma) \\
    A_{\times, n} &=2x_{b_n}\cos\iota\cos2\gamma\\
    B_{\times, n} &=2x_{a_n}\cos\iota\sin2\gamma\\
    \alpha_{M,n} &= \tan ^{-1} \left(-\frac{A_{M,n}}{B_{M,n}}\right)
\end{align}
where $M$ represents two polarization mode ($+,\times$).

\subsection{Pulasr Timing Residuals}

If GWs pass between the Earth and pulsars, the propagation path of pulses is changed, and the arrival time of pulses is also changed. The difference between the actual and predicted arrival time of pulses is called a timing residual. The timing residual induced by GWs for $a$-th pulsar is written by
\begin{align}
    R_a(t, \hat{\Omega}) = \int_0^t dt\, z_a(t, \hat{\Omega}),
    \label{eq:Residual1}
\end{align}
where $z_a(t, \hat{\Omega})$ is the rate of change in the arrival time of pulses. Using the direction of unit vector $\hat{p}_a=(\sin\theta_a\cos\phi_a,\sin\theta_a\sin\phi_a,\cos\theta_a)$, $z_a$ can be written as follows:
\begin{align}
     z_a(t,\hat{\Omega}) = \frac{1}{2}\frac{\hat{p}_a^i\hat{p}_a^j}{1+\hat{p}_a\cdot\hat{\Omega}}\left(h_{ij}(t,\hat{\Omega})-h_{ij}(t_p, \hat{\Omega})\right),
     \label{eq:residual}
\end{align}
where $t_p=t-L_a\left(1+\hat{p}_a\cdot\hat{\Omega}\right)$ is time when the GW passes the $a$-th pulsar and $L_a$ is the distance from it to the Earth. In Eq. (\ref{eq:residual}) the first and second terms are called "the Earth term" and "the pulsar term", respectively. When the GW wavelength is much shorter than the typical pulsar distance ($\sim \mathrm{kpc}$), i.e, GW frequency is much larger than $10^{-13}$ Hz, the pulsar term contributes as random noise with zero mean. In this work, we consider a situation where the GW frequensy is $\gtrsim 10^{-11}\,\mathrm{Hz}$ and, therefore, we consider only the Earth term in the following section.

Using the antenna beam pattern $F_a^M(\hat{\Omega})$ given by \citet{Anholm2009}, Eq. (\ref{eq:Residual1}) is written as follows:
\begin{align}
    R_a(t, \hat{\Omega}) &= \sum_{M=+,\times}F_a^M\int_0^t dt h_M(t), \label{eq:Residua2}\\
    F_a^M(\hat{\Omega}) &=\frac{1}{2}\frac{\hat{p}_a^i\hat{p}_a^i}{1+\hat{p}_a\cdot\hat{\Omega}}e_{ij}^M(\hat{\Omega}).
    \label{eq:F}
\end{align}
Then we assume pulsar distribution as uniform in the sky, and average Eq. (\ref{eq:Residua2}) with the direction of pulsars. Eq. (\ref{eq:Residua2}) depends the direction of pulsars only through $F_a^M(\hat{\Omega})$ but this averages out to zero. Then we use the root mean square of $F_a^M(\hat{\Omega})$ which is constant. Furthermore, the dependence of polarization vanishes by this procedure. Then we can calculate the GW amplitude with either polarization. Therefore the averaged timing residual can be written as
\begin{align}
	R_M(t) &= \bar{F}\int_0^t dt ~ h_M(t), \\
	\bar{F} &= \int d\hat{p}_a^3 \sqrt{\left(F_a^M\right)^2}
\end{align}
where $\hat{F}$ is the root mean square of $F_a^M(\hat{\Omega})$. Substituting Eq. (\ref{eq:GWsHarmonics}), we obtain
\begin{align}
    R_M(t) \propto h_0 \sum_n \frac{1}{nf}\sqrt{A_{M,n}^2 + B_{M,n}^2}\sin(2\pi nft + nl_0 + \alpha_{M,n}) \label{eq:Residual3}
\end{align}

\section{Upper limits on ultra-low-frequency GWs from eccentric binary}
\label{section3}

In this section, we develop a formalism to derive upper limits on ultra-low-frequency GWs from an eccentric SMBH binary. The signal-to-noise ratio of PTA satisfies the following equation (see \citetalias{MCJ15}):
\begin{align}
    \rho^2 &= \sum_b \sum_{a>b}\frac{8}{T}\int df \frac{|R_a(f)|^2|R_b(f)|^2}{S_{n,a}^2}, \label{eq:sn1}\\
    S_{n,a}&=\delta t_a\sigma_a^2,
\end{align}
where $T$ and $1/\delta t_a$ are the observing time span and cadence, respectively, and $\sigma_a$ is the root mean square in the timing residuals for the $a$-th pulsar. Here, we consider sky averaged timing residuals and Eq. (\ref{eq:sn1}) is written as
\begin{align}
    \rho^2 = \sum_{M=+,\times}\frac{1}{2}N_p(N_p-1)\frac{8}{T}\int df\frac{|R_M(f)|^4}{\delta t^2\sigma^4}, \label{eq:sn2}
\end{align}
where $N_p$ is the number of pulsars in PTAs. 
Considering Parseval's theorem to change the frequency integral to a time integral, Eq. (\ref{eq:sn2}) can be written approximately as
\begin{align}
    \rho^2 \approx \sum_{M=+,\times}\frac{1}{2}N_p(N_p-1)T\int_0 ^T dt \frac{|R_M(t)|^4}{\delta t^2 \sigma^4} .
\end{align}
Substituting Eq. (\ref{eq:Residual3}), we obtain
\begin{align}
\label{eq:integral}
    \rho^2 \propto &h_0^4\sum_{M=+,\times}\int_0^T dt \nonumber\\
    &\times \left( \sum_n \frac{1}{nf} \sqrt{A_{M,n}^2 + B_{M,n}^2} \sin(2\pi nft + nl_0 + \alpha_{M,n}) \right)^4.
\end{align}
In the high frequency limit ($ft \gg 1$), right hand side can be approximated as $h_0^4(\eta_+ + \eta_\times)/f^4$, where $\eta$ is the factor which depend on orbital elements  (see Appendix \ref{AppA}). Therefore, upper limits in the high frequency limits behave as
\begin{align}
    h_{\mathrm{lim}}^{\mathrm{HIGH}} \propto f (\eta_+ +\eta_\times)^{-1/4}. \label{eq:high_limits}
\end{align}
On the other hand, in the low-frequency limit ($ft \ll 1$), the sine function is expanded as a power series
\begin{align}
    \sin(2\pi nft + \phi_n) =& \sin\phi_n + (2\pi nft)\cos\phi_n - \frac{(2\pi nft)^2}{2!}\sin\phi_n \nonumber \\ &- \frac{(2\pi nft)^3}{3!}\cos\phi_n + \mathcal{O}(f^4t^4).
\end{align}
The first term in this expansion degenerate with the distance to the pulsar. The second and third terms degenerate with the pulse period and spin-down rate respectively. Therefore, these terms are absorbed when parameter fitting of the pulsar model is carried out. Consequently, upper limits in the low-frequency limits is obtained from the fourth term and behave as,
\begin{align}
    h_{\mathrm{lim}}^{\mathrm{LOW}} &\propto f^{-2} (\xi_++\xi_\times)^{-1}, \label{eq:low_limits}\\
    \xi_M &= \left|\sum_n n^2\sqrt{A_{M,n}^2 + B_{M,n}^2}\cos(nl_0 + \alpha_{M,n})\right|. \label{eq:xi}
\end{align}

Coefficients on the right hand side of Eqs. (\ref{eq:high_limits}) and (\ref{eq:low_limits}) can be given by current PTAs observation. The most recent limits on GWs from individual SMBH binary comes from NANOGrav \citep{NANOGrav11year}, which placed 95\% upper limits with $f_{\mathrm{gw}}=8\,\mathrm{nHz}$ as a function of sky position from an analysis of their 11-year data set (see Figure 5 in their paper). 
Then we can set upper limits on the GW from eccentric SMBH binary at higher and lower frequencies as follows:
\begin{align}
 	h_{\mathrm{lim}} =& h_{\mathrm{lim}}^{\mathrm{NANOGrav}} \left( f_{\mathrm{gw}} = 8 \,\mathrm{nHz},\, \hat{\Omega} \right)\nonumber \\ 
 	&\times \left[\left(\frac{8\,\mathrm{nHz}}{2f}\right)^2(\xi_++\xi_\times)^{-1} + \left(\frac{2f}{8 \,\mathrm{nHz}}\right)(\eta_++\eta_\times)^{-1} \right]
    \label{eq:upper_limits}
\end{align}
Note that $f_{\mathrm{gw}}$ is the frequency of GWs from a circular binary and corresponding to $n=2$. Therefore we chose $2f$ as the normalized frequency in Eq. (\ref{eq:upper_limits}). The right-hand side of Eq. (\ref{eq:upper_limits}) is determined by giving orbital parameters of the assumed SMBH binary $(m_1,\,m_2,\,a,\,e,\,l_0,\,\iota,\,\gamma)$. Finally, we obtain constraints on these parameters by comparing $h_{\mathrm{lim}}$ and $h_0$.

\section{Application}
\label{section4}

\begin{figure*}
	\includegraphics[width=18cm]{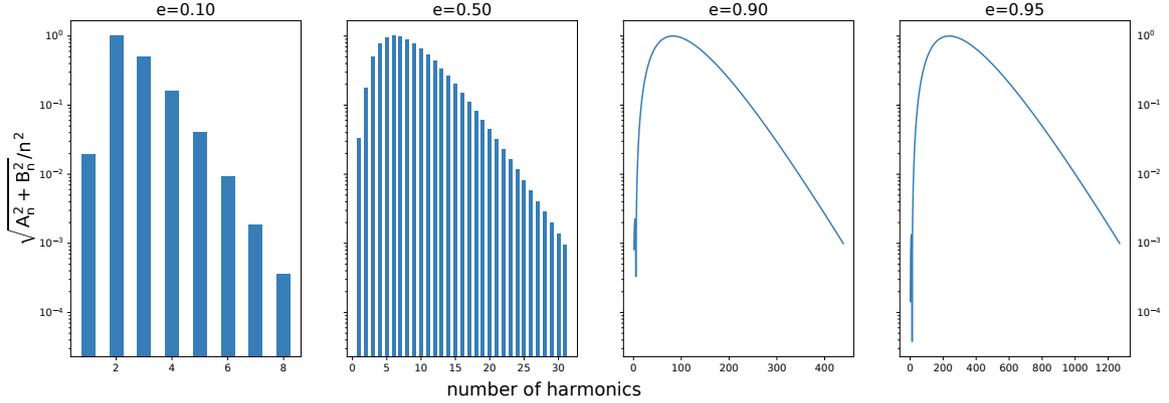}
    \caption{The contribution of higher harmonic components to $\xi_+$ normalized by the most contributing harmonic component for $(l_0,\,\iota,\,\gamma) = (0^\circ,0^\circ,0^\circ)$.     }
    \label{fig:amplitude}
\end{figure*}

In this section, we apply our formalism to several nearby SMBH binary candidates. In the numerical evaluation of upper limits, it is necessary to terminate the calculation of the sum of $\eta$ and $\xi$ with the required accuracy. In our work, we terminate the calculation when the following conditions are satisfied:
\begin{align}
    \frac{\frac{1}{i}\sqrt{A_{M,i}^2 + B_{M,i}^2} }{ \mathrm{max}_n \left(\frac{1}{n}\sqrt{A_{M,n}^2 + B_{M,n}^2}\right)} < 10^{-3} \quad (\mathrm{for} \,\eta), \\
    \frac{i^2\sqrt{A_{M,i}^2 + B_{M,i}^2}}{\mathrm{max}_n \left(n^2\sqrt{A_{M,n}^2 + B_{M,n}^2}\right)} < 10^{-3} \quad (\mathrm{for}\, \xi).
\end{align}

In Fig.~\ref{fig:amplitude}, we show the contribution of higher harmonic components to $\xi_+$ for several values of eccentricity. In this figure, other binary parameters are set as $(l_0,\,\iota,\,\gamma) = (0^\circ,0^\circ,0^\circ)$. We can see that the contribution of higher harmonics is larger for a larger value of eccentricity. For example, $n \sim 300$ modes are contributing the most for the case of $e = 0.95$. In this case, we need to conduct the summation of Eq.~(\ref{eq:xi}) up to $n = 1270$, while the summation up to $n = 8$ is sufficient for $e = 0.1$.

\begin{figure*}
	\includegraphics[width=18cm]{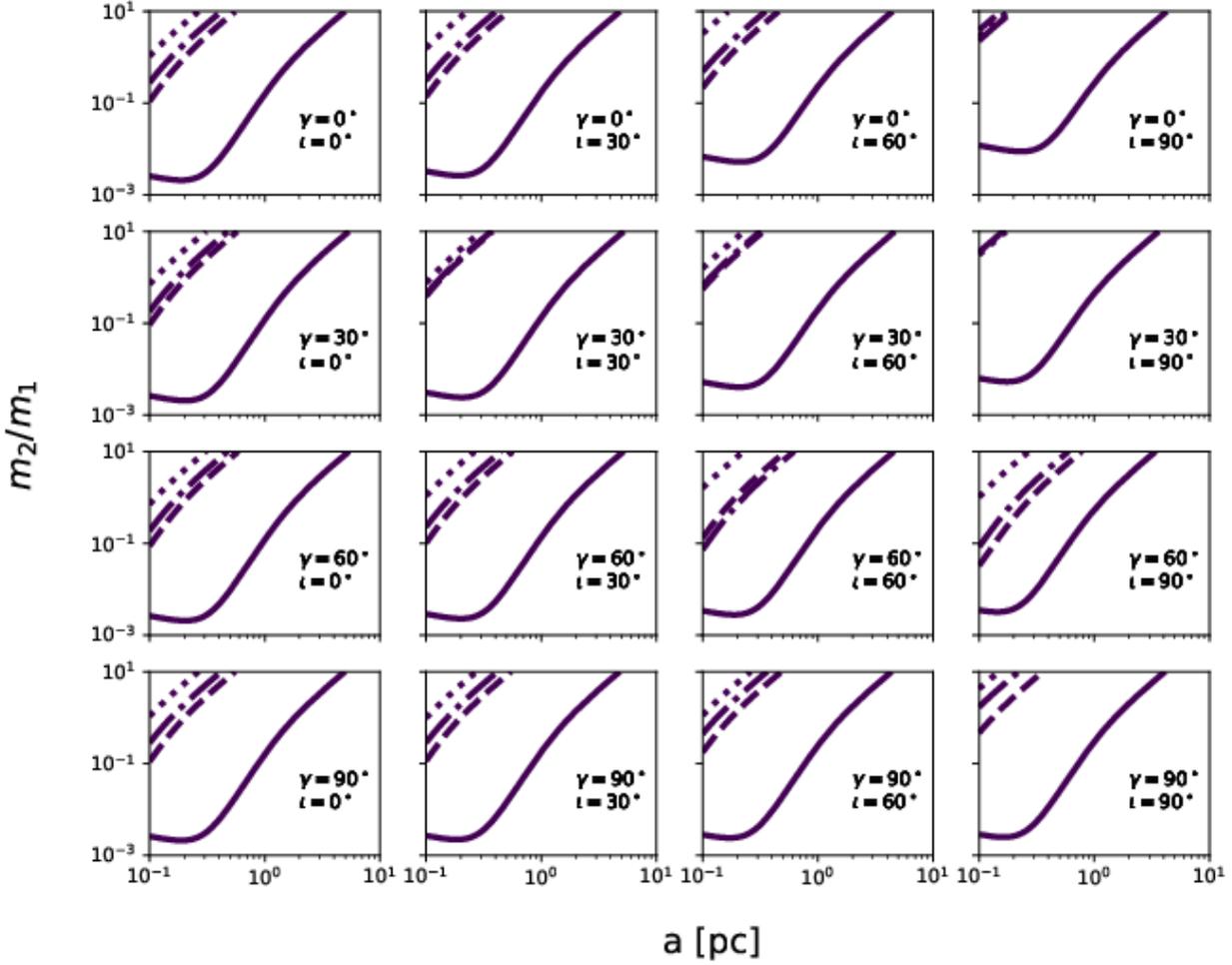}
    \caption{The rejected parameter space of an eccentric SMBH binary in the center of M87 for $e=0.9$. Solid, dashed, dot-dashed, dot lines means points of $h_0=h_{\mathrm{lim}}$ for $l_0 = 0^{\circ},60^{\circ},120^{\circ},180^{\circ}$, respectively. Regions above each curve is rejected.}
    \label{fig:rot}
\end{figure*}

First, let us show limits on a possible SMBH binary located at the center of M87 suggested by \cite{Lena}. The mass of the SMBH in the center of M87 is estimated to be $6.5 \times 10^{9}M_{\odot}$ and the distance from earth is $16.8\,\mathrm{Mpc}$ \citep{EHT}. The value of NANOGrav's limit in the direction of M87 is approximately given as:
\begin{align}
	h_{\mathrm{lim, M87}}^{\mathrm{NANOGrav}} \left( f_{\mathrm{gw}} = 8 \,\mathrm{nHz},\, \hat{\Omega}_{\mathrm{M87}} \right)\approx 3.66 \times 10^{-15}.
\end{align}
Fig. \ref{fig:rot} represents the rejected parameter space of a possible eccentric SMBH binary in the center of M87 for $e=0.9$. Solid, dashed, dot-dashed, dot lines represent the boundary of $h_0=h_{\mathrm{lim}}$ for $l_0 = 0^{\circ},60^{\circ},120^{\circ},180^{\circ}$, respectively.  The value of $h_0$ is greater than $h_{\mathrm{lim}}$ in the region below each curve and, therefore, the corresponding parameter sets are rejected. The limit becomes stronger as $l_0$ decreases. This is because small $l_0$ corresponds to a binary which starts near the pericenter and consequently the GW amplitude becomes stronger. The constraint curves do not vary significantly with the value of $(\iota, \gamma)$, although a smaller inclination angle leads to slightly stronger constraint. These parameters affect the relative power of two polarizations ($+,\times$), but do not the total energy of emitted GWs. Thus, hereafter, we fix $(\iota, \gamma)$ to $(0, 0)$.

In the case of $l_0 = 0^{\circ}$, the mass ratio is strongly constrained especially for $a \lesssim 0.3~{\mathrm{pc}}$: typically $m_2/m_1 \lesssim 3 \times 10^{-3}$. On the other hand, the lower limit on the semi-major axis is a function of the mass ratio for $m_2/m_1 \gtrsim 3 \times 10^{-3}$ and roughly given as $a \gtrsim 2 (m_2/m_1)^{0.3}~{\mathrm{pc}}$. The constraints on the semi-major axis is weaker by about one order in the case of $l_0 = 60^{\circ}$ and even slightly weaker for $l_0 = 120^{\circ}$ and $180^{\circ}$.

\begin{figure}
	\includegraphics[width=\columnwidth]{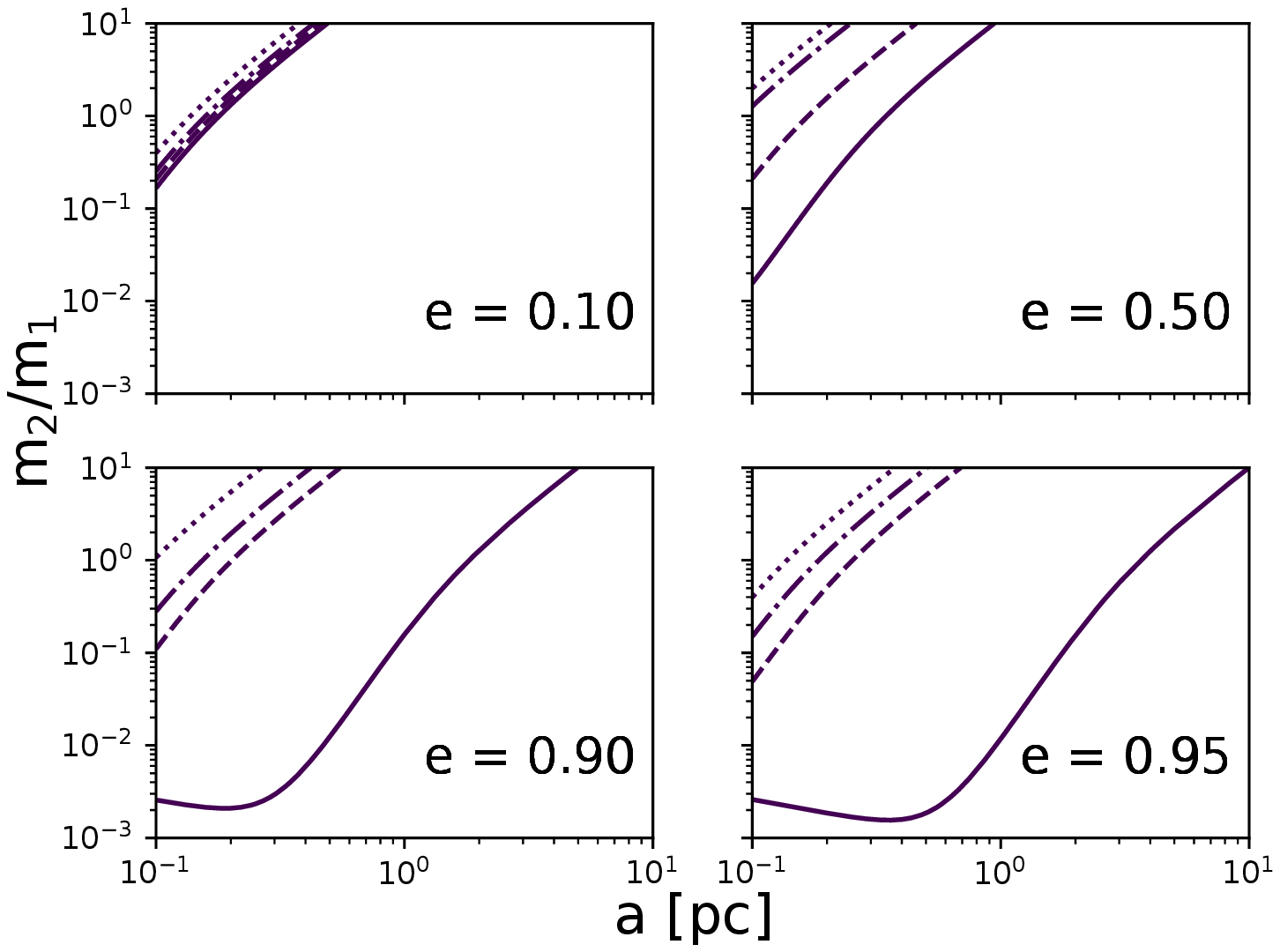}
    \caption{The rejected parameter space for $e = 0.1, 0.5, 0.9$ and $0.95$ fixing $(\iota, \gamma)=(0,0)$. The line types are the same as Fig.~\ref{fig:rot}.}
    \label{fig:e}
\end{figure}

Next, in Fig.~\ref{fig:e}, we show the rejected parameter space for different values of eccentricity fixing $(\iota, \gamma)=(0,0)$. The constraints drastically change with eccentricity in the case with $l_0 = 0^{\circ}$, while the change is not significant for other values of $l_0$. This is because the binary separation changes relatively rapidly for $l_0 = 0^{\circ}$ (pericenter). In fact, in the case with $l_0 = 0^{\circ}$, the constraints on semi-major axis at $m_2/m_1 = 0.1$ improve by a factor of 5 and 2 for the change of eccentricity from $0.5$ to $0.9$ and from $0.9$ to $0.95$, respectively.

Here it should be noted that the change of the constraint curve is not monotonic with the change of eccentricity for $l_0 = 180^{\circ}$ (apocenter). This is because there are two competing factors that affect the GW amplitude from a binary at apocenter. The first is that higher eccentricity leads to a larger separation between two SMBHs, which weakens the GW amplitude. The second is that the shape of the binary orbit near the apocenter becomes sharper for large eccentricity, which enhances the GW amplitude. Therefore, we consider that the former effect is more effective than the latter for $e=0.5$ and, conversely the latter effect becomes relatively more effective for $e=0.1$ and $e=0.9$.

For a high eccentricity binary at pericenter $l_0 = 0^{\circ}$, we can see a turnover in the curve as a function of $a$. This turnover can be interpreted as follows. From Eqs.~(\ref{eq:f}), (\ref{eq:h0}) and (\ref{eq:upper_limits}) in low-frequency cases ($2f \ll 8\, \mathrm{nHz}$), i.e. $a \gtrsim 1 \,\mathrm{pc}$, $h_0/h_{\mathrm{lim}}$ behave as:
\begin{align}
    \label{eq:behavior}
	\frac{h_0}{h_{\mathrm{lim}}} \propto \frac{m_2 f^2}{a} \propto \frac{(q+1)q}{a^4},
\end{align}
where $q = m_2/m_1$ is mass ratio. Because limit curves correspond to $h_0/h_{\mathrm{lim}}=1$, the relation between $a$ and $q$ is as follows:
\begin{align}
    \log(1 + q) + \log q - 4\log a + C = 0,
\end{align}
where $C$ is the coefficient of the right hand side in Eq.~(\ref{eq:behavior}). For large and smalls values of $q$, the relation is simplified to,
\begin{align}
    \label{eq:lowFreq}
    \log q = 
    \begin{cases}
        2\log a - C/2 \quad &(q \gg 1) \\
        4\log a - C \quad & (q \ll 1).
    \end{cases}
\end{align}
For this reason, the slope of the curves slightly vary in $q \sim 1$. On the other hand, in high frequency cases ($2f \gg 8\,\mathrm{nHz}$), i.e. $a \lesssim 0.1 \,\mathrm{pc}$, we have,
\begin{align}
    \frac{h_0}{h_{\mathrm{lim}}} \propto \frac{m_2}{af} \propto \sqrt{\frac{qa}{1+q}}.
\end{align}
For $q \ll 1$, the relation reduces to,
\begin{align}
    \label{eq:highFreq}
    \log q = - \log a - 2 C'.
\end{align}
Considering Eqs.~(\ref{eq:lowFreq}) and (\ref{eq:highFreq}), we can understand that there is a turnover at $a \sim 0.3~{\mathrm pc}$. 

We also apply our formalism to other galaxies. In \cite{NANOGrav2021}, NANOGrav applied their 95\% upper limits on GW amplitudes from single sources in galaxies listed in 2MASS Redshift Survey \citep{2MRS}. They calculated signal-to-noise ratio of GWs from these galaxies assuming they have an equal-mass SMBH binary in the center. These galaxies were sorted in descending order with respect to signal-to-noise ratio. We derive constraints for five galaxies with largest signal-to-noise ratios. Table \ref{tab:2MASS_Catalogue} is a list of five galaxies considered here: the SMBH mass, the distance from earth and NANOGrav's 95\% upper limtis $h_{\mathrm{lim}}^{\mathrm{NANOGrav}}(f_{\mathrm{gw}}=8\,\mathrm{nHz})$. In Table \ref{tab:e05} to \ref{tab:e095}, we list upper limits on mass ratio of hypothetical SMBH binaries in these galaxies for $e=0.5,\,0.9$ and $0.95$, and $l_0 = 0^\circ, \,60^\circ,\,120^\circ$ and $180^\circ$, fixing $a = 1~{\mathrm{pc}}$.

\begin{table}
    \centering
    \caption{The SMBH mass, the distance from earth and NANOGrav's 95\% upper limtis of five galaxies considered here \citep{NANOGrav2021}.}
    \label{tab:2MASS_Catalogue}
    \begin{tabular}{cccc}
        \hline
        2MASS Name & Mass & Dist & $h_{\mathrm{lim}}^{\mathrm{NANOGrav}}$ \\
         & [$\log(m_1/M_{\bigodot})$] & [Mpc] &\\
        \hline \hline
        J13000809+2758372 & 10.32 & 112.2 &  $3.17\times 10^{-15}$\\
		J12304942+1223279 & 9.82 & 16.8 & $3.66 \times 10^{-15}$\\
		J04313985-0505099 & 10.23 & 63.8 & $1.04 \times 10^{-14}$\\
		J12434000+1133093 & 9.67 & 18.6 & $3.77 \times 10^{-15}$\\
		J13182362-3527311 & 9.89 & 53.4 & $2.94\times 10^{-15}$\\
		\hline
    \end{tabular}
\end{table}

\begin{table}
	\centering
	\caption{The mass-ratio upper limits as a function of $l_0$ with $a = 1\,\mathrm{pc}$ and $e = 0.5$.}
	\label{tab:e05}
	\begin{tabular}{ccccc} 
		\hline
		\multicolumn{1}{c}{2MASS Name} & \multicolumn{4}{c}{Mass ratio $m_2/m_1$} \\
		 &  $l_0 = 0^\circ$ & $l_0 = 60^\circ$ & $l_0 = 120^\circ$ & $l_0 = 180^\circ$
		\\
		\hline \hline
		J13000809+2758372 & 4.79 & 22.9 & 75.9 & 110\\
		J12304942+1223279 & 12.0 & 52.5 & 174 & 251 \\
		J04313985-0505099 & 9.12 & 39.8 & 145 & 209 \\
		J12434000+1133093 & 20.9 & 91.2 & 331 & 479\\
		J13182363-3527311 & 14.5 & 63.1 & 229 & 331\\
		\hline
	\end{tabular}
\end{table}

\begin{table}
	\centering
	\caption{The mass-ratio upper limits as a function of $l_0$ with $a = 1\,\mathrm{pc}$ and $e = 0.9$.}
	\label{tab:e09}
	\begin{tabular}{ccccc} 
		\hline
		\multicolumn{1}{c}{2MASS Name} & \multicolumn{4}{c}{Mass ratio $m_2/m_1$}\\
		 & $l_0 = 0^\circ$ & $l_0 = 60^\circ$ & $l_0 = 120^\circ$ & $l_0 = 180^\circ$
		\\
		\hline \hline
		J13000809+2758372 & 0.033 & 14.5 & 25.1 & 63.1\\
		J12304942+1223279 & 0.158 & 36.3 & 63.1 & 158\\
		J04313985-0505099 & 0.110 & 27.5 & 47.9 & 120\\
		J12434000+1133093 & 0.437 & 63.1 & 110 & 275\\
		J13182363-3527311 & 0.251 & 43.7 & 75.9 & 191\\
		\hline
	\end{tabular}
\end{table}

\begin{table}
	\centering
	\caption{The mass-ratio upper limits as a function of $l_0$ with $a = 1\,\mathrm{pc}$ and $e = 0.95$.}
	\label{tab:e095}
	\begin{tabular}{ccccc} 
		\hline
		\multicolumn{1}{c}{2MASS Name} & \multicolumn{4}{c}{Mass ratio $m_2/m_1$}\\
		 & $l_0 = 0^\circ$ & $l_0 = 60^\circ$ & $l_0 = 120^\circ$ & $l_0 = 180^\circ$
		\\
		\hline \hline
		J13000809+2758372 & 0.00302 & 9.12 & 17.4 & 33.1\\
		J12304942+1223279 & 0.0120 & 22.9 & 43.7 & 75.9\\
		J04313985-0505099 & 0.0100 & 17.3 & 33.1 & 63.1\\
		J12434000+1133093 & 0.0363 & 39.8 & 75.9 & 145 \\
		J13182363-3527311 & 0.0191 & 27.5 & 52.5 & 100\\
		\hline
	\end{tabular}
\end{table}

\section{Summary and discussion}
\label{section5}

In this paper, we developed a formalism for constraining ultra-low frequency GWs from a SMBH binary with eccentric orbit. Following \citetalias{MCJ15}, we calculated signal-to-noise ratio of GWs by Taylor expanding the waveform and using the third-order term that is not absorbed by fitting pulsar parameters. Furthermore, using upper limits on GWs from single sources at $8~{\mathrm{nHz}}$ obtained by NANOGrav's 11-year data set, we derived constraints on binary parameters of a hypothetical SMBH binary in the center of M87. We found that the constraints depend strongly on the orbital eccentricity and initial phase while they do not depend significantly on the inclination and the azimuthal angle of pericenter. The obtained upper limits on mass ratio are typically $(m_2/m_1) < 0.16$ for $e = 0.9,\, a= 1\,\mathrm{pc}$ for pericenter ($l_0=0^\circ$). We also applied our formalism to several other SMBHs in nearby massive galaxies probed by NANOGrav.

In our calculation, we assumed a uniform distribution of MSPs in the sky. In fact, MSPs used in PTA experiments have a non-uniform distribution and many of them are located within the Galactic plane. Although the anisotropy of MSP distribution will not change the frequency dependence of GW constraints, it will affect the normalization. It is expected that GW constraints would become stronger (weaker) for a sky region with more (less) MSPs. Quantitative discussion with numerical integration of the factor in Eq.~(\ref{eq:F}) is beyond the scope of the current paper and will be presented elsewhere.

We also assumed the binary orbit does not change in the observing time span, which is typically $\sim$ 10 years, because we mainly consider the ultra-low frequency range. However, a binary orbit with semi-major axis of $\lesssim 0.1~{\mathrm{pc}}$ will shrink and can become circular by the GW emission with such a time scale. Therefore, if the GW waveform and frequency change, our formalism may not be valid.

\section*{Acknowledgements}

SH is supported by JSPS KAKENHI Grant Number 20J20509. KT is partially supported by JSPS KAKENHI Grant Numbers 15H05896, 16H05999, 17H01110, and 20H00180, Bilateral Joint Research Projects of JSPS, and the ISM Cooperative Research Program (2020-ISMCRP-2017).

\section*{Data Availability}

The data used in calculation for upper limits on mass ratio of 2MASS galaxies are available from this link \url{https://github.com/nanograv/nanograv_galaxy_catalog_2MRS}. 



\bibliographystyle{mnras}
\bibliography{mnras_template} 



\onecolumn
\appendix



\section{High Frequency Integral}

\label{AppA}
In \citetalias{MCJ15}, the integration of the $\sin^4$ term at high frequencies $(ft \gg 1)$ is
\begin{align}
	\int_0^T dt \sin^4(2\pi ft + \phi) = \frac{3}{8}T - \frac{1}{4}\sin(2T+2\phi) +\frac{1}{32}\sin(4T + 4\phi).
\end{align}
The first term of the right hand side is $O(T)$ and the second and third terms are $O(1)$. Then the second and third term could be neglected. In our work, we need to calculate the integral of Eq.~(\ref{eq:integral}):
\begin{align}
	&\int_0^T dt \left(\sum_n\frac{1}{n}\sqrt{A_{M,n}^2 + B_{M,n}^2}\sin(2\pi nft + nl_0 + \alpha_{M,n})\right)^4 \nonumber \\
    &= \sum_{i,j,k,l} A_{M,ijkl}\int_0^T \sin(2\pi ift + \phi_{M,i}) \sin(2\pi jft + \phi_{M,j})\sin(2\pi kft + \phi_{M,k}) \sin(2\pi lft +\phi_{M,l})\,dt,
\end{align}
where $A_{M,ijkl}$ is a unified term of coefficients of each $\sin$ and $\phi_{M,n}=nl_0+\alpha_{M,n}$. Using formula of trigonometric function, we transform the integrand:
\begin{align}
	\label{eq:sin_to_cos}
	&\quad \sin(2\pi ift + \phi_{M,i}) \sin(2\pi jft + \phi_{M,j}) \sin(2\pi kft + \phi_{M,k}) \sin(2\pi lft +\phi_{M,l}) \nonumber \\
    &= \frac{1}{4}\left[\cos \{2\pi (i-j)ft + \phi_{M,i}-\phi_{M,j}\} - \cos \{2\pi (i+j)ft + \phi_{M,i}+\phi_{M,j}\}\right] \nonumber \\
    &\quad \times \left[\cos \{2\pi (k-l)ft +\phi_{M,k}-\phi_{M,l}\} -\cos \{2\pi (k+l) ft +\phi_{M,k} +\phi_{M,l}\}\right]
\end{align}
By expanding the right hand side, we obtain four $\cos \times \cos$ terms. One of them can be transformed as follows:
\begin{align}
	\label{eq:cos_to_cos}
	&\quad\cos \{2\pi (i-j)ft + \phi_{M,i}-\phi_{M,j}\}\cos \{2\pi (k-l)ft +\phi_{M,k}-\phi_{M,l}\}  \nonumber \\
    &=\frac{1}{2}\left[\cos\{2\pi (i-j+k-l)ft +\phi_{M,i}-\phi_{M,j}+\phi_{M,k}-\phi_{M,l}\}+\cos\{2\pi (i-j-k+l)ft +\phi_{M,i}-\phi_{M,j}-\phi_{M,k}+\phi_{M,l}\}\right].
\end{align}
If $\,i-j+k-l=0\,$, the time dependence of the first term of the right hand side is vanished. Then, this term contributes to the signal-to-noise ratio at $O(T)$ as a consequence of time integration. On the other hand, if $\,i-j+k-l\neq 0\,$, the time dependence of this term remains and this term behave $O(1)$ after time integration. Therefore, among the terms expressed by expanding the Eq. (\ref{eq:sin_to_cos}) and transforming it like Eq. (\ref{eq:cos_to_cos}), only the terms whose the time dependence is vanished for a certain combination of $(i,j,k,l)$ has a non-negligible value after the time integration. The conditions of $(i,j,k,l)$ are the following equations:
\begin{align}
	i-j+k+l=0, \\
    i-j-k+l=0, \\
    i-j-k-l =0, \\
    i-j+k-l =0, \\
    i+j+k-l =0, \\
    i+j-k-l =0, \\
    i+j-k+l=0.
\end{align}
Writing these conditions with $f_m(i,j,k,l)=0\,(m = 1,\cdots, 7)$ from the top to the bottom, sets of $(i,j,k,l)$ satisfying $f_m=0$ can be written as follows:
\begin{align}
	\Lambda_m = \{(i,j,k,l)\in\mathbb{N}_+^4 | f_m(i,j,k,l)=0\},
\end{align}
where $\mathbb{N_+}$ is the set of positive integer. We write the sum of $\phi_{M,i},\phi_{M,j},\phi_{M,k},\phi_{M,l}$ added with the sign same as $(i,j,k,l)$ appered in $f_m$ (for example, $\Phi_{M,ijkl}^1=\phi_{M,i}-\phi_{M,j}+\phi_{M,k}+\phi_{M,l}$). Then the integration of Eq. (\ref{eq:integral}) approximate as:
\begin{align}
	&\int_0^T dt \left(\sum_n\frac{1}{n}\sqrt{A_{M,n}^2 + B_{M,n}^2}\sin(2\pi nft + nl_0 + \alpha_{M,n})\right)^4
    \approx\frac{T}{8}\sum_{m=1}^7\sum_{(i,j,k,l)\in\Lambda_m}(-1)^{m}A_{M,ijkl}\cos\Phi_{M,ijkl}^m.
\end{align}
Therefore, we obtain upper limits in Eq. (\ref{eq:upper_limits}) defining $\eta_M$ as follows:
\begin{align}
    \eta_M = \sum_{m=1}^7\sum_{(i,j,k,l)\in\Lambda_m}(-1)^{m}A_{M,ijkl}\cos\Phi_{M,ijkl}^m.   
\end{align}
In this work, the contribution of $\eta_M$ in Eq (\ref{eq:upper_limits}) is small because we consider ultra-low frequency GWs ($\le \mathrm{nHz}$).


\bsp	
\label{lastpage}
\end{document}